\documentclass[12pt,preprint]{aastex}

\newcommand{\myemail}{loris@physast.uga.edu}

\slugcomment{Submitted to: {\it The Astronomical Journal}}

\shorttitle{CH Along the Galactic Plane}
\shortauthors{Magnani, Lugo, and Dame}

\begin{document}

\title{CH 3 GHz Observations of Molecular Clouds Along the Galactic Plane}

\author{Loris Magnani and Samantha Lugo}
\affil{Department of Physics and Astronomy, The University of Georgia,
    Athens, GA 30602 \\ \myemail}

\and 

\author {T. M. Dame}
\affil{Harvard-Smithsonian Center for Astrophysics, 
60 Garden St., MS 72, Cambridge, MA 02138}

\begin{abstract}

Spectra in the CH $^2\Pi_{1/2}$, J=1/2, F=1-1 
transition at 3335 MHz were obtained in three 5-point crosses centered
on the Galactic plane at $\ell =$ 50$\arcdeg$, 100$\arcdeg$, and 110$\arcdeg$.  
The lines of sight traverse both Giant Molecular 
Clouds (GMCs) and local, smaller entities.  This transition is a good
tracer of low-density molecular gas and the line profiles are very similar to CO(1-0) 
data at nearly the same resolution.  In addition, the CH 3335 MHz line can be used 
to calibrate the CO-H$_2$ conversion factor (X$_{\rm CO}$) in low-density 
molecular gas.  Although this technique underestimates X$_{\rm CO}$ in 
GMCs, our results are within a factor of two of X$_{\rm CO}$ values
calibrated for GMCs by other techniques.  The similarity of CH and CO line profiles, 
and that of X$_{\rm CO}$ values derived from CH and more traditional techniques, 
implies that most of the molecular gas along the 
observed lines of sight is at relatively low densities ($n \le$ 10$^3$ cm$^{-3}$).
 
\end{abstract}


\section{Introduction}

The CH ground state, hyperfine, main line  
($^2\Pi_{1/2}$, J = 1/2, F = 1-1) at 3335 MHz is a good, though seldom used,
tracer of low-density molecular gas ($n \sim$ 10$^2$ - 10$^3$ cm$^{-3}$ -
Magnani et al. 2003 and references therein).
One of the earliest surveys of the CH 3 GHz lines\footnote{The $^2\Pi_{1/2}$, J = 1/2
lambda-doubling transition consists of 3 hyperfine structure components: 
The F = 1-1 main line at 3335 MHz, and the 
F = 0-1 and 1-0 satellite lines at 3264 and 3349 MHz, respectively.}
by Rydbeck et al. (1976) detected emission from all the denser types of molecular
clouds: Giant Molecular Clouds (GMCs), dark clouds, and what today would be
referred to as translucent clouds. 
The CH 3 GHz lines are 
nearly ubiquitous in molecular gas detectable via the CO(1-0) line.  
Unfortunately, the CH lines tend to be very weak, with peak antenna temperatures that 
seldom surpass 0.5 K and, more often, are in the tens of milliKelvin range.  This is partly a 
consequence of the relatively low CH abundance (compared to CO) in interstellar clouds as determined 
from observations (4 $\times$ 10$^{-8}$ with respect to H$_2$; 
e.g., Liszt \& Lucas (2002) and reference therein), and partly because of the peculiar
non-thermal excitation of the 3 GHz transitions (e.g., Rydbeck et al. 1976; Liszt \& Lucas 2002).  
However, the linear relationship of N(CH) with N(H$_2$) - at least
for N(H$_2$) $<$ 5 $\times$ 10$^{21}$ cm$^{-2}$ (Federman 1982; 
Danks, Federman, \& Lambert 1984; Mattila 1986; Rachford et 
al. 2002, Liszt \& Lucas 2002; Magnani et al. 2003; Weselak et al. 2004) - makes CH a valid tracer
of molecular gas in all but the densest cloud regions (which tend to occupy little
cloud volume - see below) and,  because the transition is optically thin in the 
interstellar medium, many of the difficulties inherent in interpreting CO(1-0) data are 
avoided.

The only systematic CH survey of the Galactic plane was made by Johansson (1979) 
and his results, though intriguing, are not well-known because they were published in 
a report of the Research Laboratory of Electronics and Onsala Space Observatory.
Basically, the Galactic plane was surveyed in the CH 3335 MHz line at 2.5$\arcdeg$
or 5$\arcdeg$ intervals from $\ell =$ 10$\arcdeg$ - 230$\arcdeg$. Like the CO emission,
the CH peaks at a galactocentric distance of 5.5 kpc (assuming R$_\odot$ = 10
kpc) and the equivalent of
the Molecular Ring (e.g., Scoville \& Sanders 1987; Dame, Hartmann, \& Thaddeus 2001;
Simon et al. 2001) is clearly seen in the CH data.  However, in comparison with 
to the CO distribution, the CH data show significant excess emission
between 7 - 10 kpc. Given that the CH distribution in this portion of the Galaxy
appeared to be intermediate between the CO and HI distributions, 
Johansson (1979) speculated that 
a portion of the CH emission arose from a low-density molecular gas regime that was not
readily traceable by CO.  The existence of such a regime has sometimes been 
proposed to explain infrared 
excess regions (e.g., Reach, Wall, \& Odegard 1998; Blitz, Bazell, \& D\'esert 1990).

The question of whether the CH 3335 MHz line can trace low-density H$_2$
not detectable in the CO(1-0) transition has been explored in some detail by
Magnani \& Onello (1993; 1995) and revisited by Magnani et al. (2003).  The consensus
opinion seems to be that deep CO spectra will generally reveal the same lines 
as seen in deep CH spectra.  The CO surveys 
used by Johansson (1979) may not have been sufficiently sensitive to
pick up the CO equivalent of the excess CH emission between 7-10 kpc.  
However, regardless of the ultimate resolution of the excess CH issue, the results of Johansson (1979)
merit further consideration.

In addition to confirming the Johansson (1979) results, CH observations of
the 3335 MHz line can be used to calibrate the CO-H$_2$ conversion factor,
X$_{\rm CO}$ [$\equiv$ N(H$_2$)/W$_{\rm CO}$, where W$_{\rm CO}$ is the 
velocity-integrated CO(1-0) antenna temperature].
Magnani \& Onello (1995) find that this method 
works well for translucent
molecular gas but is thought to break down for dark molecular clouds and
for the denser portions of GMCs.  The principal reason is that in dense
(and thus opaque) molecular regions, CH is converted into more complex
molecular species so that its abundance decreases rapidly with increasing
N(H$_2$), $n$, or opacity (e.g., Federman 1982; 
Viala 1986; Lee, Bettens, \& Herbst 1996; Liszt \& Lucas 2002).
In addition to chemistry considerations,
the presence of significant  background continuum radiation (from HII regions embedded in
GMCs) can also drive down the CH antenna temperature (Rydbeck et al. 1976).

Observations compiled from several sources by Mattila (1986) confirm this result; in 
using N(CH) as a molecular tracer for 6 GMC lines of sight underestimates N(H$_2$) 
significantly (see Figure 10 of his paper).
However, even if CH completely disappeared at high densities, GMCs are not
uniformly at $n >$ 10$^3$ cm$^{-3}$.  The molecular envelopes around dense, 
star-forming cores have low enough densities and constitute enough of 
the cloud  to drive the
average molecular volume density of GMCs down to 50 cm$^{-3}$
(Blitz 1991).  Lada, Bally, \& Stark (1991) estimate from CO observations 
that the dense cores (i.e., those with $n \ge$ 10$^4$ cm$^{-3}$)
in the L1630 (Orion B) GMC constitute only a small fraction (not greater than 19\%) 
of the total cloud mass.  The fraction of a GMC's mass and volume as a function of $n$
 likely varies greatly from cloud to cloud and has
not been adequately studied.  Nevertheless, GMC envelopes around dense cores are ideal
sources for the CH 3 GHz lines and the early CH surveys confirmed this [Rydbeck
et al. (1976); Hjalmarson et al. (1977); Genzel et al. (1979)]\footnote{In many
star-forming regions, the CH 3264 MHz satellite line is stronger than the
main line at 3335 MHz because of far-infrared pumping from HII regions -
see Rydbeck et al. (1976).}, although the CH lines were often weaker 
than those observed in dark clouds.  
However, because of the decreased abundance of CH at high molecular densities, the
CH 3335 MHz line will systematically underestimate N(H$_2$) and, thus, X$_{\rm CO}$ in GMCs.
The key issue is to determine how significant is this effect.  

In order to examine these issues further, we conducted a preliminary CH Galactic plane survey
with the NRAO
140 ft telescope, observing CH emission
from 4 regions along the Galactic plane.  Three of the regions were centered at
$b =$ 0$\arcdeg$ and $\ell =$ 50$\arcdeg$, 100$\arcdeg$, and 110$\arcdeg$,
and are described in this paper;  the other region covered the Galactic Center and,
given its uniqueness and exclusion from the Johansson
(1979) survey, the observations are described elsewhere (Magnani et al. 2005).
In \S 2 of this paper, we describe the set-up for the CH 3335 MHz and C$^{18}$O 109.8 GHz
observations, and the results
are presented in \S 3.  In \S 4, we discuss the CH - H$_2$ relation at higher
N(H$_2$) than usual and, in \S 5, derive the values of X$_{\rm CO}$ for the
various components along the observed lines of sight.  A brief conclusion closes the paper.

\section{Observations}

The CH 3335 line was observed in 3 general directions along the Galactic plane 
in 1999 March using the now-defunct  140 ft telescope in Green Bank, West Virginia. 
At 3.3 GHz the beam size of the 140 ft was 9$\arcmin$.   The observing
configuration consisted of a front end with a corrugated dual-hybrid
mode feed in which two linear polarizations were fed into a dual-channel
FET amplifier receiver.  The system temperature on the sky
ranged from 35 - 50 K, depending on the atmospheric conditions and the 
antenna elevation.  The autocorrelator was configured into two sections
of 512 channels, with each section covering a bandwidth of 5 MHz at a
velocity resolution of 0.88 km s$^{-1}$ per channel.  The total velocity
coverage of each spectrum was $\sim$ 450 km s$^{-1}$ centered on
the v$_{\rm LSR}$ velocity corresponding to the bulk of the CO emission as determined
from the Dame, Hartmann, \& Thaddeus (2001) survey.

A 5-point cross was made at each of the 3 locations with the central spectrum 
at $b =$ 0$\arcdeg$ and $\ell=$ 50$\arcdeg$, 100$\arcdeg$, and
110$\arcdeg$, and the other 4 spectra offset by 0.125$\arcdeg$ in latitude or 
longitude. Each position was observed
in ON-OFF mode with one hour total on-source integrations.  The OFF positions
were typically 10$\arcdeg$ away from the plane in $b$.  For each line of sight, 
the two polarizations were added together and the resulting spectrum was
baselined with a polynomial of order 6 and Hann smoothed
to yield an {\it rms} noise level of $\sim$ 5-8 mK.
Individual reduced spectra for the 5 positions centered on 
 ($\ell, b$) =  (50$\arcdeg$, 0$\arcdeg$) are shown in Figures 1$a$ - 1$e$ 
along with the corresponding CO(1-0) spectra from the Dame, Hartmann, \&
Thaddeus (2001) Galactic survey.  The CO data are at comparable angular and 
velocity resolution (8.8$\arcmin$ and 0.65 or 1.3 km s$^{-1}$, respectively).
For ($\ell, b$) =  (110$\arcdeg$, 0$\arcdeg$), only the central position is shown
in both CH and CO in Figure 2 (there is little change in the spectra from one position 
to another 
in this region), and for ($\ell, b$) = (100$\arcdeg$, 0$\arcdeg$), both the CH and 
CO signals were relatively weak so that all 5 spectra are summed 
together to make a single composite composite spectrum shown in Figure 3.

The three central positions ($\ell =$ 50$\arcdeg$, 100$\arcdeg$, and 110$\arcdeg$ at
$b =$ 0$\arcdeg$) were also observed in the C$^{18}$O(1-0) line at
109.8 GHz with the CfA 1.2 meter telescope, the same instrument used for the Dame et
al. (2001) CO survey.  Flat spectral baselines were obtained by position switching
every 15 s between the source position and two  emission-free positions chosen to
straddle the source in elevation. Residual baseline offsets ($<$ 1 K) were removed
by fitting straight lines to the baseline region.  At 109.8 GHz, the 256 channel
filterbank spectrometer provided a velocity resolution of 0.68 km s$^{-1}$ over
a range of 174 km s$^{-1}$.  Total integration times per position ranged from 1
to 3 hours, yielding rms noise levels of 12 - 16 mK per channel.

\section{Results}

The CH and CO profiles shown in Figures 1-3 are strikingly similar to each
other.  Everywhere there is CO(1-0) emission there is also CH emission.
However, the integration times are very different.   Each CH spectrum was
observed for at least one-hour on source, whereas the CO spectra were only
observed for a few minutes. 
Although the CH spectra presented in this paper have significantly 
higher signal-to-noise
ratios than the Johansson (1979) results (5-8 mK vs. 20 mK),  nevertheless, the
spectra for equivalent lines of sight look quite similar (the resolution of the
data in this paper is 9$\arcmin$ vs. 15$\arcmin$ for the Johansson data).

All the spectra centered on ($\ell, b$) = (50$\arcdeg$, 0$\arcdeg$) 
and the composite spectrum at ($\ell, b$) = (100$\arcdeg$, 0$\arcdeg$)
show two or more components whose velocity was assigned using the
antenna temperature-weighted velocity over the extent of each spectral
feature. We treat each velocity component as a separate spectral feature. 
In table 1, columns 1 and 2 list the observed position, column 3 shows
the antenna temperature-weighted velocity of the component,  and column
4 the velocity-integrated CH antenna temperature (defined to be
W$_{\rm CH}$) for each of the 21 velocity 
components.  From W$_{\rm CH}$, the column density can be obtained 
in the traditional manner (see Rydbeck et al. 1976) using the formula:

\begin{equation}
\label{ }
N(CH) = 2.82 \times 10^{11} [\eta_f \eta_b (T_{ex} - T_{bg})]^{-1} T_{ex} \  W_{\rm CH}  \ \
\ \ \ cm^{-2}
\end{equation} 

\noindent
where T$_{ex}$ and T$_{bg}$ are the excitation and background temperatures respectively,
$\eta_f$ is the filling fraction of CH in the telescope beam (assumed to be 1 for the
observations described in this paper), $\eta_b$ is the beam efficiency (0.79 at
3335 MHz for the 140-ft telescope - Ron Maddalena, private communication), and W$_{\rm CH}$
is measured in mK km s$^{-1}$.
Because T$_{ex}$ is usually negative and $\vert$ T$_{ex} \vert \gg$ T$_{bg}$
 for the 3 GHz lines (Genzel et al. 1979; Hjalmarson et al. 
1977; Rydbeck et al. 1976), the quantity in square brackets is nearly always close to unity (however,
see discussion in Magnani \& Onello 1995 and Liszt \& Lucas 2002).
Consequently,  N(CH) is directly
proportional to W$_{\rm CH}$ and is tabulated in column 5 of Table 1.  

Once N(CH) has been obtained, the conversion to N(H$_2$)  
follows from the empirical relation established by Mattila (1986):

\begin{equation}
\label{ }
N(H_2) = (2.1 \times 10^7)N(CH) + 2.2 \times 10^{20} \ \
\ \ \ cm^{-2}
\end{equation}

\noindent
However, as pointed out in the \S 1, this
empirical relation is calibrated only for values of N(H$_2$) $<$ 5 $\times$ 10$^{21}$
cm$^{-2}$.  Nevertheless, the values obtained by using equation 2 are listed in 
column 6 of Table 1.

\section{The N(CH) - N(H$_2$) Relation}

As Figures 1-3 make clear, the CH 3335 MHz spectral line profiles
track the CO(1-0) emission, and thus 
H$_2$, very well; at all velocities where there is CO(1-0) 
emission, there is also CH emission  - albeit at a much lower antenna temperature.
This is confirmed by plotting W$_{\rm CH}$ vs. W$_{\rm CO}$; the linear relation that
ensues has a correlation coefficient of 0.76.  However, in order to compare our 
results with earlier work, we convert W$_{\rm CH}$ to N(CH) via equation 1, and
W$_{\rm CO}$ to N(H$_2$) using the standard Galactic plane value of X$_{\rm CO}$
(1.6 $\times$ 10$^{20}$ cm$^{-2}$ K km s$^{-1}$ - Hunter et al. 1987). 
The plot of log N(CH) vs. log N(H$_2$) 
is very similar to what was obtained by Mattila (1986) and is shown
in Figure 4.  It is clear from the figure that the lines of sight with 
the greatest N(H$_2$)
tend to deviate from the Mattila relation.  Thus, N(CH) appears to
systematically underestimate
N(H$_2$) for values $>$ 5 $\times$ 10$^{21}$ cm$^{-2}$.  However, it is 
important to stress that, in this regime, the N(H$_2$) values derived from CO observations
may have large uncertainties associated with them.  N(H$_2$) is not directly determined,
and statements of how any quantity varies with N(H$_2$) in dense molecular regions
must be viewed with some caution.

A least squares
fit to our data yields N(H$_2$) = 3.60 $\times$ 10$^7$ N(CH) - 2.36 $\times$ 
10$^{20}$, somewhat similar equation 2. 
If we fit only those   
points in our data set with N(H$_2$) $<$ 5 $\times$ 10$^{21}$, 
we obtain a best fit line 
N(H$_2$) = 2.46 $\times$ 10$^7$ N(CH) + 1.47 $\times$ 10$^{20}$, 
virtually identical to the Mattila result.

Liszt \& Lucas (2002) combine translucent and dark cloud CH and CO data from Magnani
\& Onello (1995) and new CH and CO observations to create an N(CH) vs. W$_{\rm CO}$ plot
that shows three modes of behavior: (1) A slow increase of N(CH) with W$_{\rm CO}$ is evident
for both dark and translucent gas. (2) A nearly fixed, factor of 3 offset in the best
fit lines to the dark vs. translucent data sets, with the dark data having the larger
y-intercept. (3) A scatter in N(CH) at any given value of W$_{\rm CO}$ ascribed by Liszt and
Lucas to variations in the CH excitation and the local physical conditions. 

The second point is somewhat problematic because the data for MBM16 from Magnani
\& Onello (1995) are classified by Liszt \& Lucas (2002) as belonging to the
dark cloud category.  MBM16 is a translucent cloud according to the standard 
definition of translucent: 1 $\le$ A$_V \le$ 5 magnitudes (van Dishoeck \& Black 1988).
Direct measurements of the cloud extinction by the method of star counts 
conducted by Magnani \& de Vries (1986) indicate that the average extinction 
of the cloud is only 0.9 magnitudes (A$_V$) above
the background with even the most obscured regions barely above 2
magnitudes.  Nevertheless, it is possible that the CH excitation conditions 
in MBM16 differ enough from those in ``normal" translucent gas that the
CH and CO properties of this object may be more similar to those found in
dark molecular gas.    Assuming that the bimodal distribution attributed to
the data by Liszt \& Lucas (2002) does indeed exist, and is not merely a large
scatter in the CH and CO properties of the clouds, Liszt and Lucas speculate
that  the shallowness of the slope in the diffuse gas reflects the trend that
at low extinction N(CO) $\propto$ [N(CH)]$^2$ $\propto$ [N(H$_2$)]$^2$.
In the dark gas, the same shallow slope (albeit with a y-intercept a factor of
3 higher) likely reflects the decline of the CH abundance in denser gas at higher
values of W$_{\rm CO}$.

In Figure 5, following Liszt \& Lucas (2002), we present a plot of 
N(CH) vs. W$_{\rm CO}$ for our data (star symbols). 
Also included in the plot are the data Liszt \& Lucas classify as 
arising from dark and translucent gas (plotted
as diamond symbols and plus signs, respectively).  
As is clear from the figure, the Galactic
plane data fall on and extend the relationship for the dark clouds noted by
Liszt \& Lucas (2002). Thus, the Galactic plane molecular clouds observed
in this paper are as likely underabundant in CH as the dark cloud population
selected by Liszt \& Lucas (2002).

\section{The Value of X$_{\rm CO}$ in Giant Molecular Clouds}

Magnani \& Onello (1995) describe in detail how to derive X$_{\rm CO}$ using
CH 3335 MHz and CO 115 GHz data.   In essence, equations 1 and 2 are
used to obtain N(H$_2$) from the CH data, and this result is divided by 
W$_{\rm CO}$ at similar resolution 
to immediately yield X$_{\rm CO}$.  This was done for all 21 components
in our sample and the results are tabulated in Table 2 and shown 
graphically in Figure 6.
 
The lines of sight centered on and around ($\ell, b) =$ (50$\arcdeg$, 0$\arcdeg$)
typically show 3 separate components: a relatively weak feature at $\sim$ 6 km s$^{-1}$
that likely arises from very local, small, molecular clouds, and stronger emission from
Inner Galaxy GMCs at v$_{\rm LSR} \sim$ 45 and 57 km s$^{-1}$.  This interpretation is
confirmed by Figure 7a which shows the C$^{18}$O(1-0) spectrum for 
($\ell, b) =$ (50$\arcdeg$, 0$\arcdeg$).  This transition traces dense molecular gas 
typically found in the cores of GMCs or dark clouds ($n >$ 10$^4$ cm$^{-3}$), and is
detected for this line of sight only at v$_{\rm LSR} \sim$ 45 and 57 km s$^{-1}$. 
The average X$_{\rm CO}$ for
the 6 km s$^{-1}$ gas is 1.9 $\times$ 10$^{20}$, while the Inner Galaxy X$_{\rm CO}$
values are somewhat lower; 1.2 $\times$ 10$^{20}$ and 1.3 $\times$ 10$^{20}$
for the 45 and 57 km s$^{-1}$ features, respectively.  These relatively low values
of X$_{\rm CO}$ may be the result of the CH method underestimating the
true value of N(H$_2$) in GMCs, as discussed above.  
However, if the correct value of X$_{\rm CO}$ in these Inner
Galaxy regions is  obtained from the standard 1.6 $\times$ 10$^{20}$ value used by
Hunter et al. (1997), the CH method underestimates N(H$_2$) and,
consequently, X$_{\rm CO}$, by no more than a factor of 2.

The lines of sight centered on and around ($\ell, b) =$ (100$\arcdeg$, 0$\arcdeg$)
have only one component at v$_{\rm LSR} \sim$ -50 km s$^{-1}$ that arises from molecular
clouds in the Perseus spiral arm.  The C$^{18}$O spectrum from the central position
shows relatively strong emission at v$_{\rm LSR} \sim$ -50 km s$^{-1}$ (see Figure
7b) indicating that the molecular gas along this line of sight has a dense 
component.  The values of X$_{\rm CO}$ derived using the CH
method produce values of X$_{\rm CO}$ ranging from 0.5 - 1.5 $\times$ 10$^{20}$ with
an average of 0.8 $\times$ 10$^{20}$.  Again, these determinations most likely  
underestimate the real values.  However,  as in the case of Inner Galaxy GMCs, the
underestimate is not severe; no more than a factor of 2-3.

The 5-point cross centered on $b =$ 0$\arcdeg$ and $\ell =$ 100$\arcdeg$ 
has a much lower W$_{\rm CO}$ than the other two regions, and the CH 3335 MHz line is 
also so weak that the 5 lines of
sight have to be averaged in order to identify the two components (see Figure 3).
In contrast to the preceeding two regions, the corresponding values of X$_{\rm CO}$ 
derived from the CH data are larger than
the standard values by factors of 3 and 9, seemingly implying that the CH 
method may be overestimating N(H$_2$) for these lines of sight.  However,
it is more likely given the weakness of the CO and CH lines 
in the direction  ($\ell, b) =$ (100$\arcdeg$, 0$\arcdeg$) that this emission is
not tracing 
molecular gas in GMCs or dark cloud complexes but, rather, molecular gas in 
lower density, lower 
opacity objects such as translucent clouds. In support of this conclusion,
the C$^{18}$O spectrum for the central position shows no signal to an rms
of 16 mK.
In these types of clouds, X$_{\rm CO}$ can vary by more than an
order of magnitude from cloud to cloud and even over a single cloud, 
and can take on values both substantially higher and lower than the
standard value (Magnani \& Onello 1995; Magnani et al. 1998). 

For those lines of sight with W$_{\rm CO}$ values greater than 10 
K km s$^{-1}$,
indicative of the presence of GMCs or other high column and volume density
clouds, the X$_{\rm CO}$ values derived from the CH method are systematically
lower than the conventional value for the Galactic plane by factors of 2-3.
For those lines of sight with W$_{\rm CO}$ more similar to what is expected
from somewhat lower column density 
molecular clouds - like dark clouds,
$X_{\rm CO}$ tends to be close to the conventional value.  Finally,
for the lines of sight where the two components have W$_{\rm CO}$ values 
lower than those of dark clouds, more similar to what is seen
in translucent molecular gas, the X$_{\rm CO}$ values are larger than normal; 
typical of what is
obtained when using the CH method on translucent clouds (Magnani
\& Onello 1995).

\section{Conclusions}

The CH 3335 MHz observations described in this paper confirm that this transition
is a good tracer of molecular gas at low densities.  Figures 1-3 underscore
how strikingly similar CH 3335 MHz profiles can be to CO(1-0) profiles at similar
resolution.  Although the weakness
of the line in most Galactic environments probably precludes its use in large-scale surveys, 
the result from the undersampled Galactic plane survey made by Johansson (1979) 
- that there is more molecular gas between the Molecular Ring and the Solar Circle 
than has been traced by the CO surveys - should probably be revisited.  If this gas
exists, it can undoubtedly also be traced by the CO(1-0) line, but integration times longer
than what has commonly been employed by the large-scale surveys are probably needed.

As first noted by Magnani \& Onello (1995), for lower density molecular
gas typical of that found in small, local molecular clouds, X$_{\rm CO}$ as determined
from the CH method is consistent
with other determinations - though variations from cloud to cloud can be
significant.  For GMCs, the data presented here show that N(H$_2$) and, consequently,
X$_{\rm CO}$ is systematically
underestimated by the CH method.  This is what would be expected from the chemistry
of the molecule as described in \S 1.  Surprisingly, the underestimate is not
more than a factor of 2-3 compared to what is obtained from other methods.
Given astrochemistry models of dense molecular gas environments, this result
implies that most of the volume of GMCs is not at high molecular densities
($ n >$ 10$^3$ cm$^{-3}$) 
and the lower density envelopes contribute  significantly to the overall
molecular mass (see also, Lada, Bally, \& Stark 1991; Blitz, 1991; Magnani, Hartmann,
\& Speck 1996).

The results described above are limited, covering 21 velocity components distributed over
 15 lines of sight
along 3 principal directions in the Galactic plane.  Unfortunately, the capacity to obtain
more data is even more limited.  Currently, in the Western Hemisphere, the
only radio telescope equipped to make CH 3 GHz observations is at Arecibo.  The
nature of the Arecibo dish is such that only a limited portion of the Galactic
Glane can be surveyed (the Galactic Center is unaccessible with that telescope).
Given the promising results from our preliminary survey project, we hope that in the not too
distant future the 100 meter NRAO telescope at Green Bank will be outfitted with
an upper S-band receiver so that a proper survey might be conducted.

\acknowledgements
We thank Ben Engebreth for help with some of the observations at the 140-ft telescope
and Emily Brown for help with some of the data reduction.  We also thank an anonymous
referee for comments that improved the presentation.

\clearpage

\begin{deluxetable}{rrrrrr}
\tabletypesize{\scriptsize}
\tablecaption{CH Observations and Derived Quantities}
\tablewidth{0pt}
\tablehead{
\colhead{$\ell$ } & \colhead{$b$} & \colhead{component\tablenotemark{a}}  & \colhead{W$_{\rm CH}$} &
\colhead{N(CH)\tablenotemark{b}} & \colhead{N(H$_2$)\tablenotemark{c}} \\ 
\colhead{$\arcdeg$}   & \colhead{$\arcdeg$} & \colhead{km s$^{-1}$} & \colhead{ K km s$^{-1}$} &
\colhead{ cm$^{-2}$} & \colhead{ cm$^{-2}$} 
}
\startdata
 49.875 & 0.000 &    6.2 & 0.081  & 2.89 $\times$ 10$^{13}$  &  0.83 $\times$ 10$^{21}$ \\
        &       &   46.0 & 0.382  & 1.36 $\times$ 10$^{14}$  &  3.08 $\times$ 10$^{21}$ \\
        &       &   59.3 & 0.258  & 9.21 $\times$ 10$^{13}$  &  2.15 $\times$ 10$^{21}$ \\
 50.000 & 0.125 &    6.8 & 0.161  & 5.75 $\times$ 10$^{13}$  &  1.43 $\times$ 10$^{21}$ \\
        &       &   44.6 & 0.216  & 7.71 $\times$ 10$^{13}$  &  1.84 $\times$ 10$^{21}$ \\
        &       &   55.5 & 0.263  & 9.39 $\times$ 10$^{13}$  &  2.19 $\times$ 10$^{21}$ \\ 
 50.000 & 0.000 &    6.7 & 0.135  & 4.82 $\times$ 10$^{13}$  &  1.23 $\times$ 10$^{21}$ \\
        &       &   45.4 & 0.424  & 1.51 $\times$ 10$^{14}$  &  3.39 $\times$ 10$^{21}$ \\
        &       &   57.4 & 0.387  & 1.38 $\times$ 10$^{14}$  &  3.12 $\times$ 10$^{21}$ \\
 50.000 & -0.125 &   4.9 & 0.184  & 6.57 $\times$ 10$^{13}$  &  1.60 $\times$ 10$^{21}$ \\
        &       &   47.9 & 0.717  & 2.56 $\times$ 10$^{14}$  &  5.60 $\times$ 10$^{21}$ \\
 50.125 & 0.000 &    5.9 & 0.099  & 3.53 $\times$ 10$^{13}$  &  9.61 $\times$ 10$^{20}$ \\
        &       &   43.7 & 0.349  & 1.25 $\times$ 10$^{14}$  &  2.84 $\times$ 10$^{21}$ \\
        &       &   55.0 & 0.306  & 1.09 $\times$ 10$^{14}$  &  2.51 $\times$ 10$^{21}$ \\
100.000 & 0.000 &   -5.0 & 0.139  & 4.96 $\times$ 10$^{13}$  &  1.26 $\times$ 10$^{21}$ \\
        &       &  -31.4 & 0.313  & 1.12 $\times$ 10$^{14}$  &  2.57 $\times$ 10$^{21}$ \\
109.875 & 0.000 &  -51.0 & 0.140  & 5.00 $\times$ 10$^{13}$  &  1.27 $\times$ 10$^{21}$ \\
110.000 & 0.125 &  -50.8 & 0.200  & 7.14 $\times$ 10$^{13}$  &  1.72 $\times$ 10$^{21}$ \\
110.000 & 0.000 &  -51.1 & 0.271  & 9.67 $\times$ 10$^{13}$  &  2.25 $\times$ 10$^{21}$ \\
110.000 & -0.125 & -51.6 & 0.368  & 1.31 $\times$ 10$^{14}$  &  2.97 $\times$ 10$^{21}$ \\
110.125 & 0.000 &  -52.8 & 0.342  & 1.22 $\times$ 10$^{14}$  &  2.78 $\times$ 10$^{21}$ \\
\enddata


\tablenotetext{a}{The antenna
temperature-weighted velocity is given for each spectal component (see Figures 1-3).}

\tablenotetext{b}{N(CH) is derived from the integrated antenna temperature
in column 4 after correcting for the beam efficiency, the beam filling
fraction, and assuming $\vert$ T$_{ex} \vert \gg$ T$_{bg}$. See text 
for details.}

\tablenotetext{c}{N(H$_2$) derived from N(CH) via equation 2 
(see text).}

\end{deluxetable}

\begin{deluxetable}{rrrrrrr}
\tabletypesize{\scriptsize}
\tablecaption{CH and CO Observations and Derived Quantities for GMCs}
\tablewidth{0pt}
\tablehead{
\colhead{$\ell$ } & \colhead{$b$} & \colhead{component\tablenotemark{a}}  & 
\colhead{W$_{\rm CO}$\tablenotemark{b}} & \colhead{N(H$_2$)\tablenotemark{c}} &
\colhead{f \tablenotemark{d}} & \colhead{X$_{\rm CO}$\tablenotemark{e}} \\
\colhead{$\arcdeg$}   & \colhead{$\arcdeg$} & \colhead{km s$^{-1}$}  
& \colhead{ K km s$^{-1}$} &
\colhead{ cm$^{-2}$} &  & \colhead{$\times$ 10$^{20}$}
}
\startdata
 49.875 & 0.000 &    6.2     
                &      5.43 &  0.87 $\times$ 10$^{21}$ &  0.95 & 1.5  \\
        &       &   46.0     
                &     20.66 &  3.31 $\times$ 10$^{21}$ &  0.93 & 1.5  \\
        &       &   59.3     
                &     15.26 &  2.44 $\times$ 10$^{21}$ &  0.88 & 1.4  \\
 50.000 & 0.125 &    6.8     
                &     12.49 &  2.00 $\times$ 10$^{21}$ &  0.72 & 1.1  \\
        &       &   44.6     
                &     12.24 &  1.96 $\times$ 10$^{21}$ &  0.94 & 1.5  \\
        &       &   55.5     
                &     23.21 &  3.72 $\times$ 10$^{21}$ &  0.59 & 0.9  \\
 50.000 & 0.000 &    6.7     
                &     10.97 &  1.75 $\times$ 10$^{21}$ &  0.70 & 1.1  \\
        &       &   45.4     
                &     27.20 &  4.36 $\times$ 10$^{21}$ &  0.78 & 1.2  \\
        &       &   57.4     
                &     30.09 &  4.82 $\times$ 10$^{21}$ &  0.65 & 1.0  \\
 50.000 & -0.125 &   4.9     
                &      7.07 &  1.13 $\times$ 10$^{21}$ &  1.42 & 2.3  \\
        &       &   47.9     
                &     52.38 &  8.38 $\times$ 10$^{21}$ &  0.67 & 1.1  \\
 50.125 & 0.000 &    5.9     
                &      2.88 &  4.60 $\times$ 10$^{20}$ &  2.09 & 3.3  \\
        &       &   43.7     
                &     21.70 &  3.48 $\times$ 10$^{21}$ &  0.82 & 1.3  \\
        &       &   55.0     
                &     19.07 &  3.05 $\times$ 10$^{21}$ &  0.82 & 1.3  \\
100.000 & 0.000 &   -5.0     
                &      3.18 &  5.08 $\times$ 10$^{20}$ &  2.48 & 4.0  \\
        &       &  -31.4     
                &      1.77 &  2.84 $\times$ 10$^{20}$ &  9.05 & 14.5  \\
109.875 & 0.000 &  -51.0          
                &      8.47 &  1.35 $\times$ 10$^{21}$ &  0.94 & 1.5  \\
110.000 & 0.125 &  -50.8          
                &     23.06 &  3.69 $\times$ 10$^{21}$ &  0.47 & 0.7  \\
110.000 & 0.000 &  -51.1          
                &     36.18 &  5.79 $\times$ 10$^{21}$ &  0.39 & 0.6  \\
110.000 & -0.125 & -51.6          
                &     44.09 &  7.06 $\times$ 10$^{21}$ &  0.42 & 0.7  \\
110.125 & 0.000 &  -52.8          
                &     51.65 &  8.27 $\times$ 10$^{21}$ &  0.34 & 0.5  \\
\enddata


\tablenotetext{a}{
same as Table 1}

\tablenotetext{b}{Integrated CO(1-0) line emission from the data of
Ungerechts, Hunbanhowar, \& Thaddeus (2000); Dame, Hartmann, \& Thaddeus (2001).}

\tablenotetext{c}{N(H$_2$) derived from W$_{\rm CO}$ using X$_{\rm CO} =$ 1.6 $\times$
10$^{20}$ - Hunter et al. (1997).}

\tablenotetext{d}{ratio of N(H$_2$) derived from CH
(see Table 1) to that derived from CO.}

\tablenotetext{e}{X$_{\rm CO}$ in units of cm$^{-2}$ [K km s$^{-1}$]$^{-1}$
derived from the data in column 6 of Table 1 and column 4 of this table.}

\end{deluxetable}

\clearpage

\begin{figure}
\figurenum{1a}
\plotone{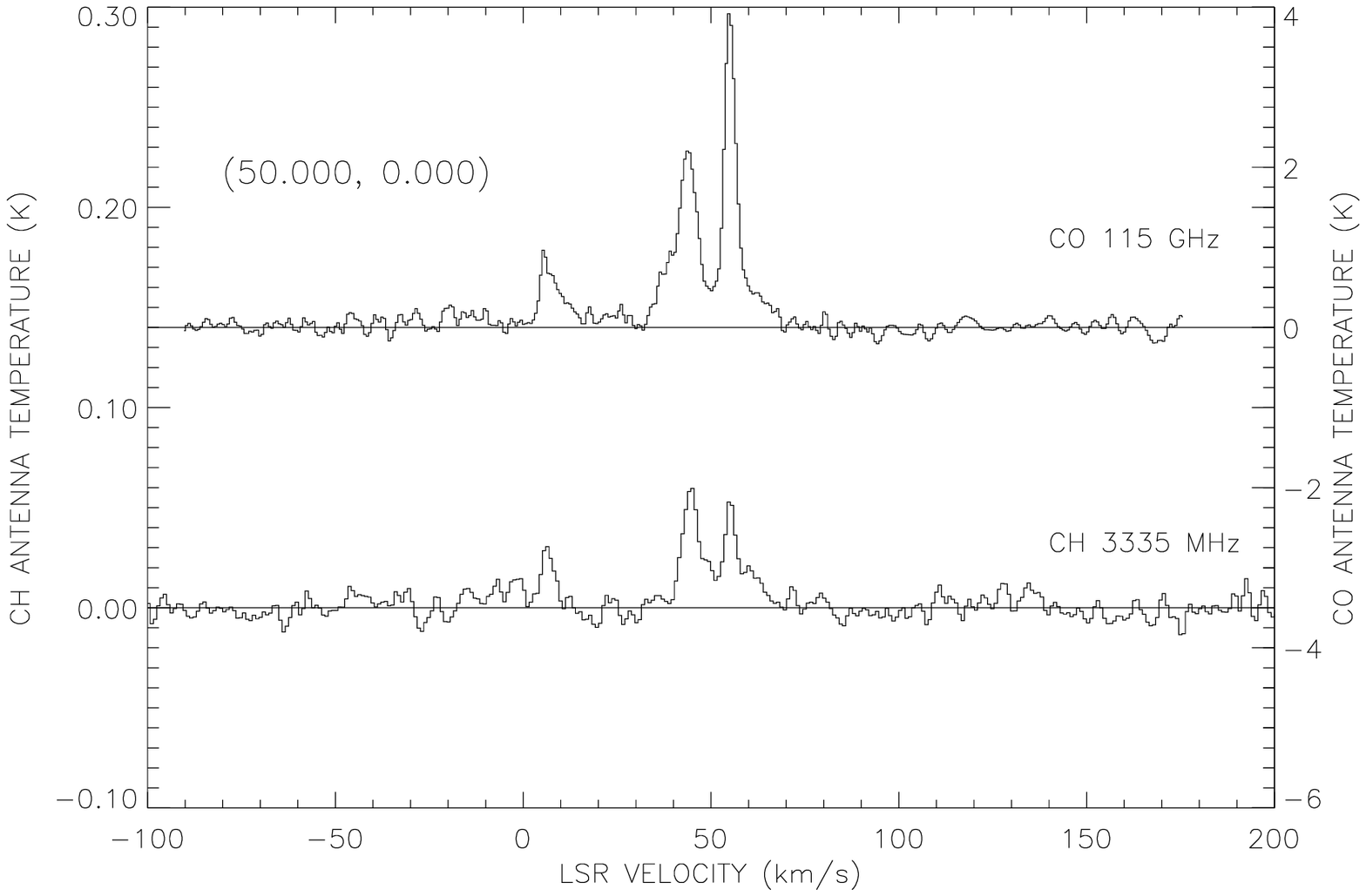}
\caption{
At bottom, spectrum of the CH $^2\Pi_{1/2}$, J=1/2, F=(1-1) transition
at 3335 MHz for $\ell =$ 50.0$\arcdeg$, $b =$ 0.0$\arcdeg$. 
The beamsize is 9$\arcmin$ and the velocity
resolution is 1.8 km s$^{-1}$ after Hann smoothing.
Above, CO(1-0) spectrum
from Ungerechts, Umbanhowar, \& Thaddeus (2000) for the same position
 with a beamsize of 8.7$\arcmin$ and a velocity resolution of 0.65 km s$^{-1}$.
}
\end{figure}

\begin{figure}
\figurenum{1b}
\plotone{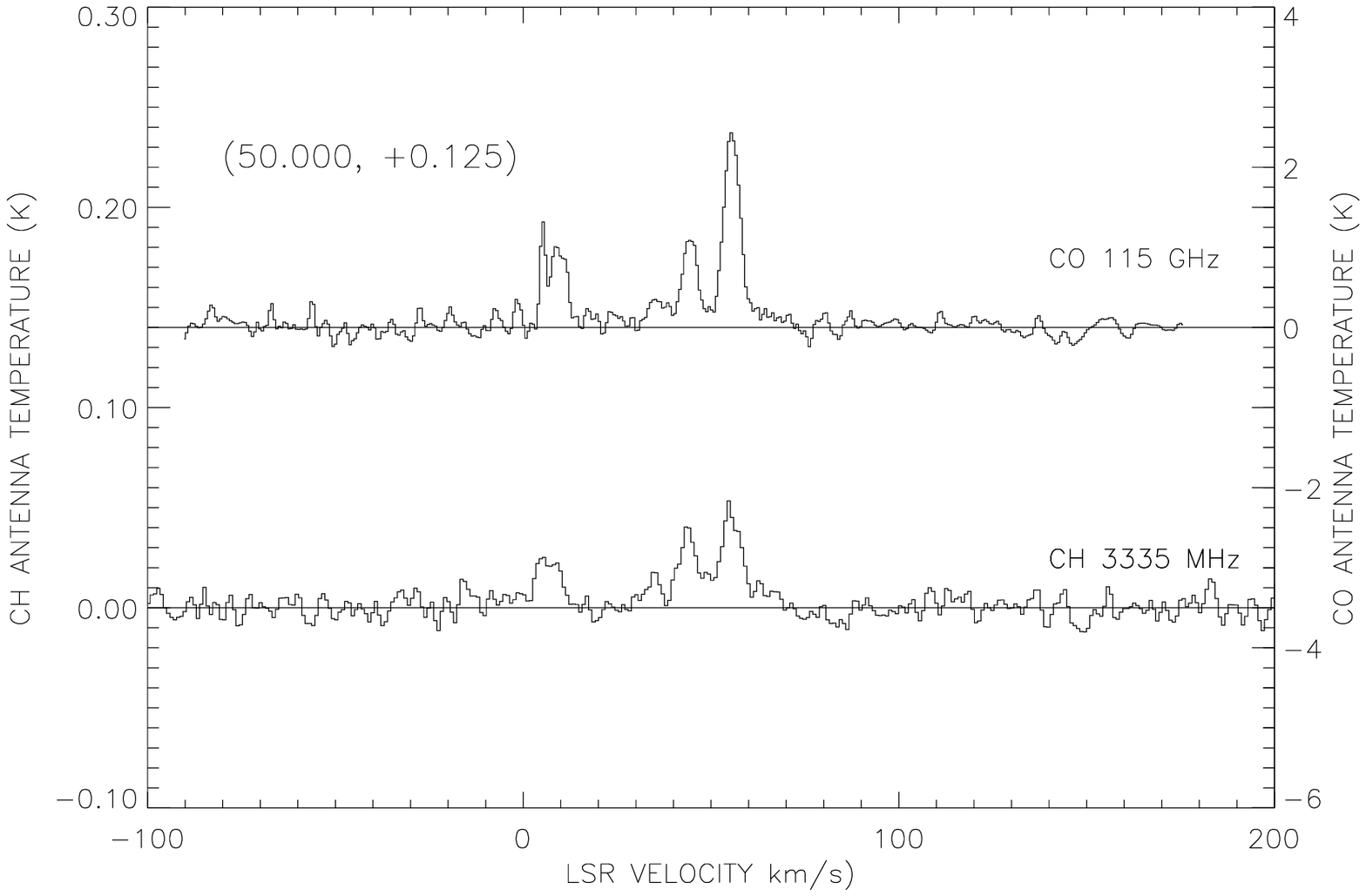}
\caption{
Same as Figure 1a, but for $\ell =$ 50.0$\arcdeg$, $b =$ 0.125$\arcdeg$. 
}
\end{figure}

\begin{figure}
\figurenum{1c}
\plotone{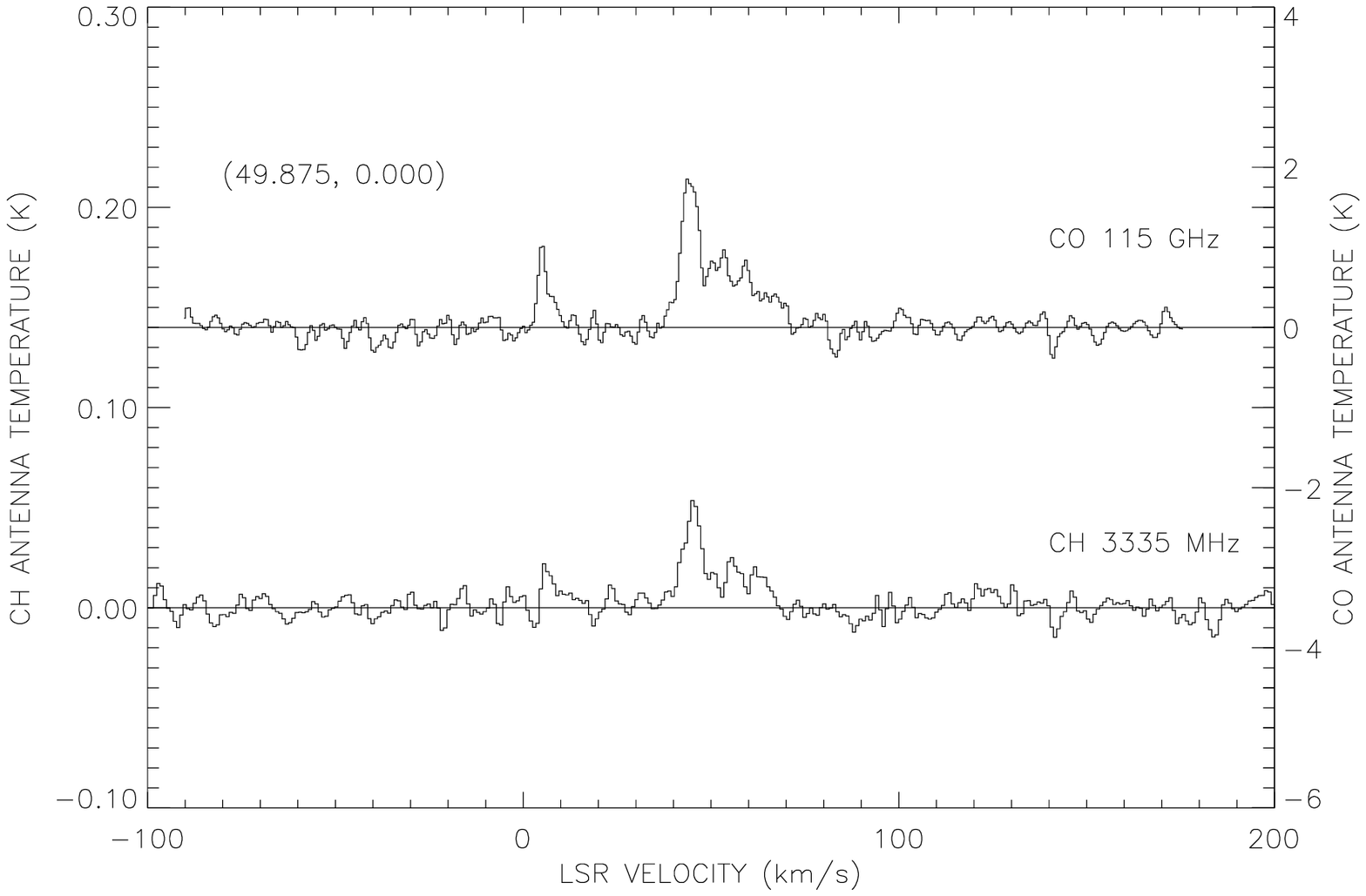}
\caption{
Same as Figure 1a, but for $\ell =$ 49.875$\arcdeg$, $b =$ 0.0$\arcdeg$. 
}
\end{figure}

\begin{figure}
\figurenum{1d}
\plotone{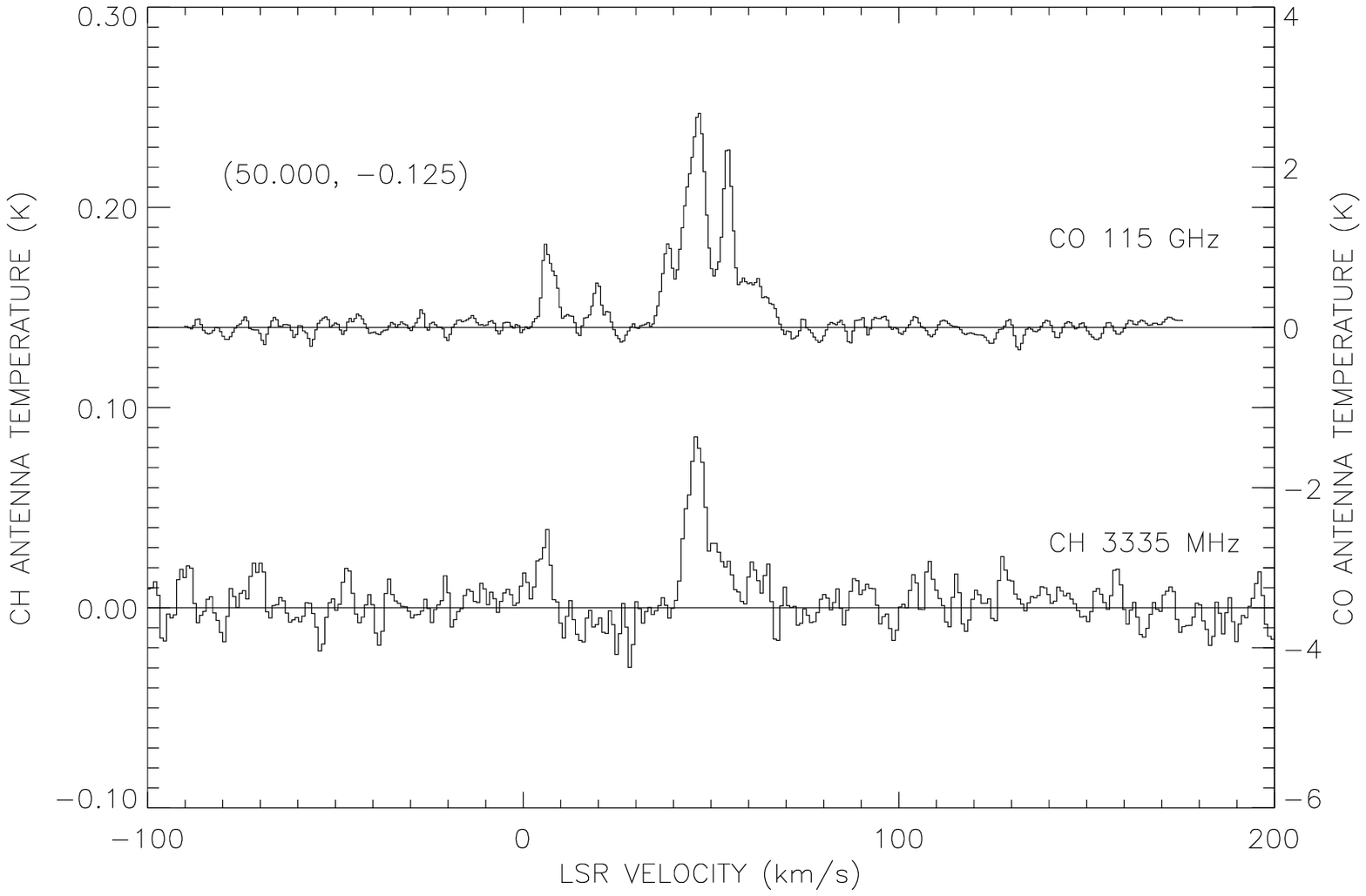}
\caption{
Same as Figure 1a, but for $\ell =$ 50.0$\arcdeg$, $b = -$0.125$\arcdeg$.
}
\end{figure}

\begin{figure}
\figurenum{1e}
\plotone{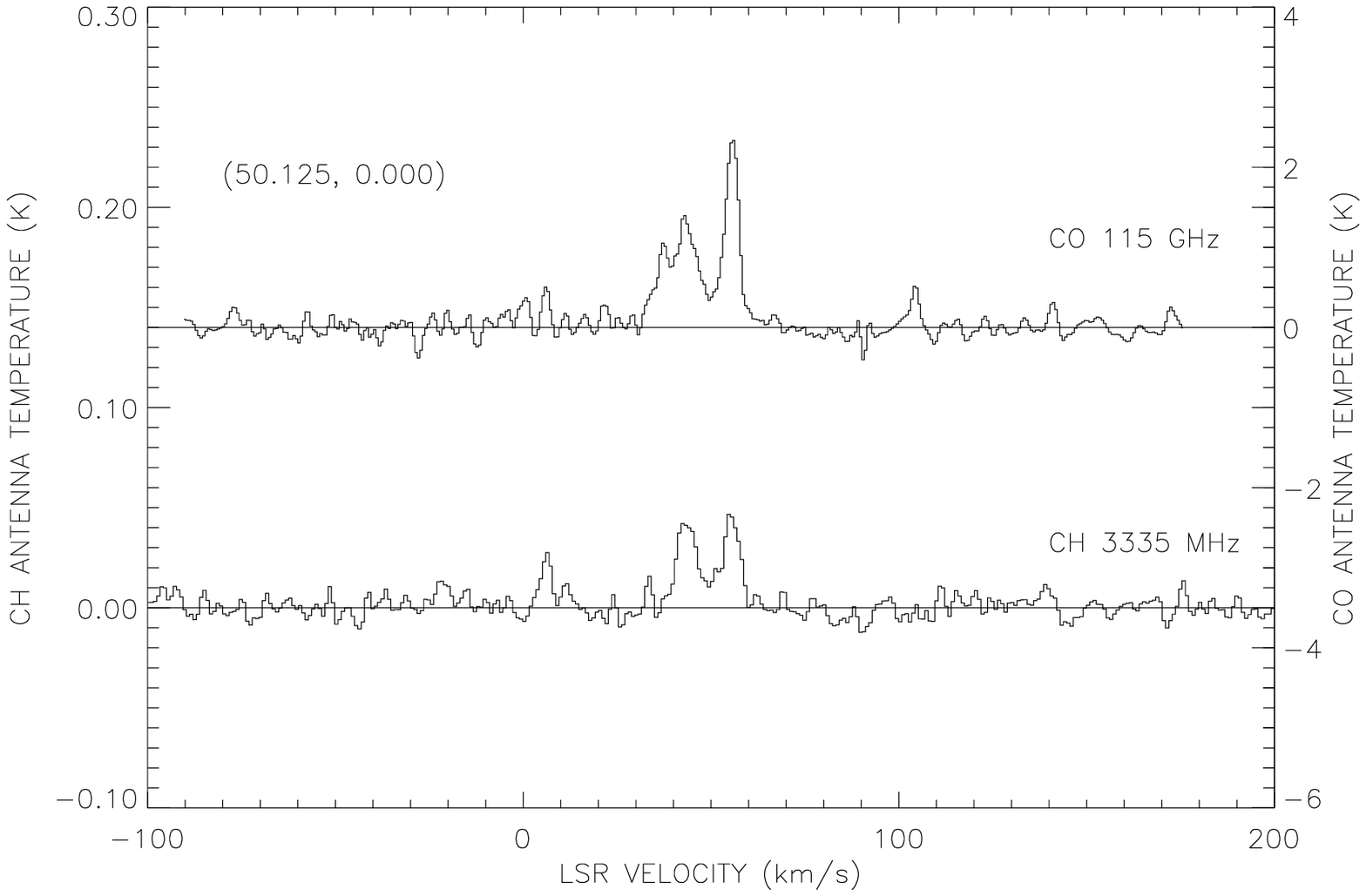}
\caption{
Same as Figure 1a, but for $\ell =$ 50.0$\arcdeg$, $b =$ 0.0$\arcdeg$.
}
\end{figure}

\begin{figure}
\figurenum{2}
\plotone{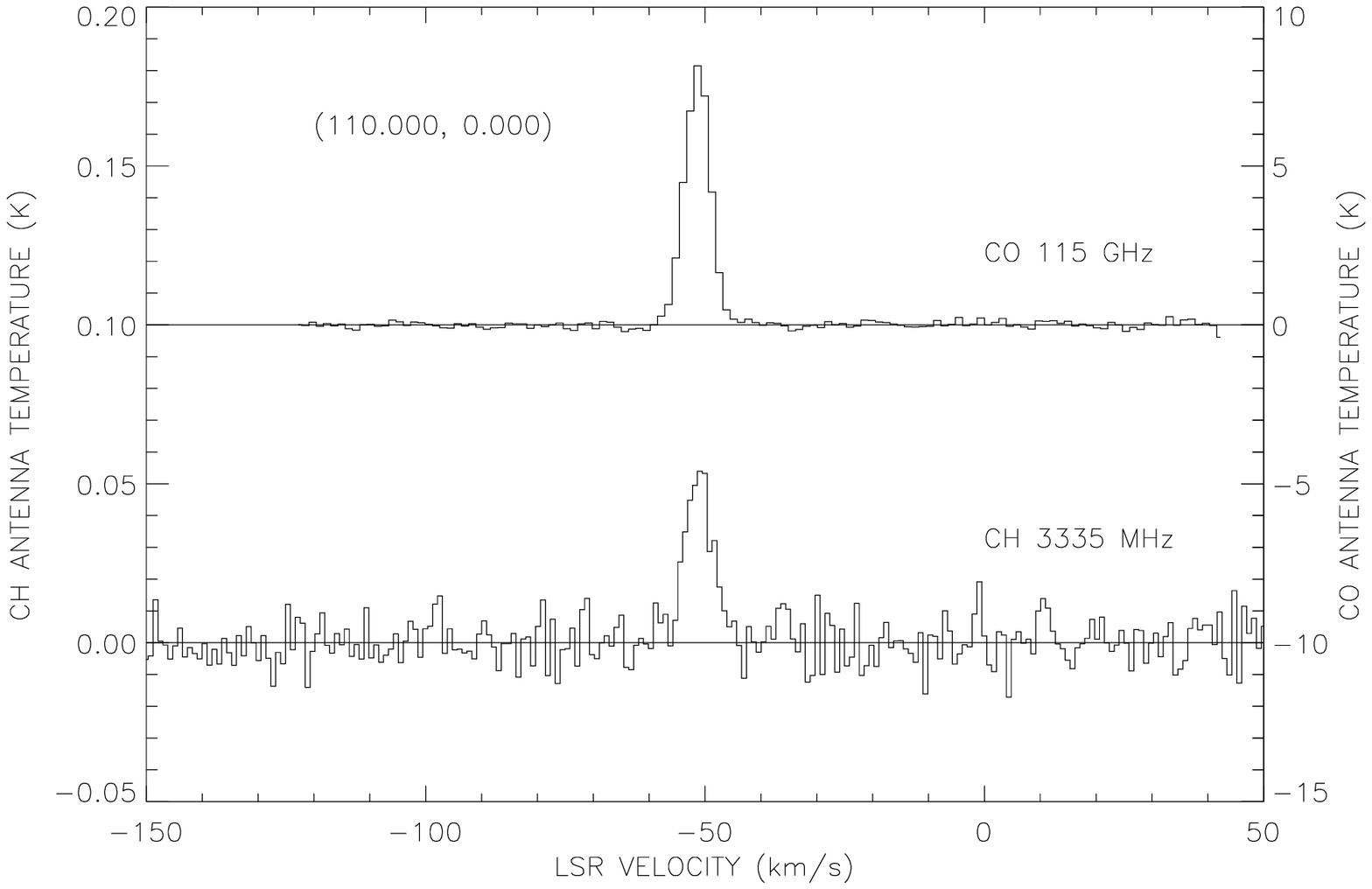}
\caption{
Same as Figure 1, but for $\ell =$ 110.0$\arcdeg$, $b =$ 0.0$\arcdeg$. 
The velocity resolution of the CO spectrum is 1.3 km s$^{-1}$.
}
\end{figure}

\begin{figure}
\figurenum{3}
\plotone{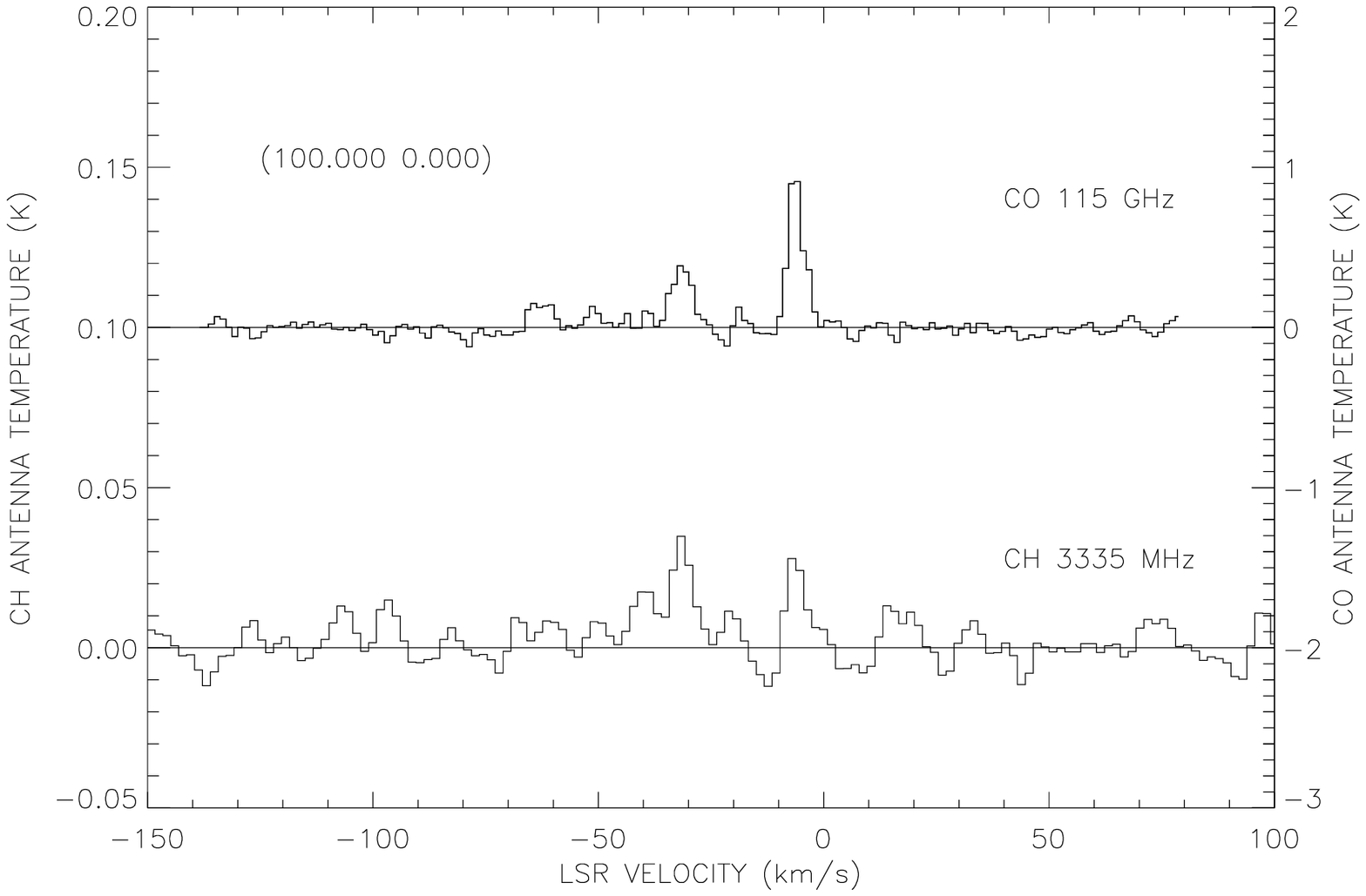}
\caption{
Same as Figures 1 and 2, but for the average of the 5 spectra centered on
$\ell =$ 100.0$\arcdeg$, $b =$ 0.0$\arcdeg$.  Both the CO and CH data are
averaged over the same 5-point cross. The CH data are Hann smoothed to a 
velocity resolution of 1.8 km s$^{-1}$.
The CO data are at a resolution of 1.3 km s$^{-1}$.
}
\end{figure}

\begin{figure}
\figurenum{4}
\plotone{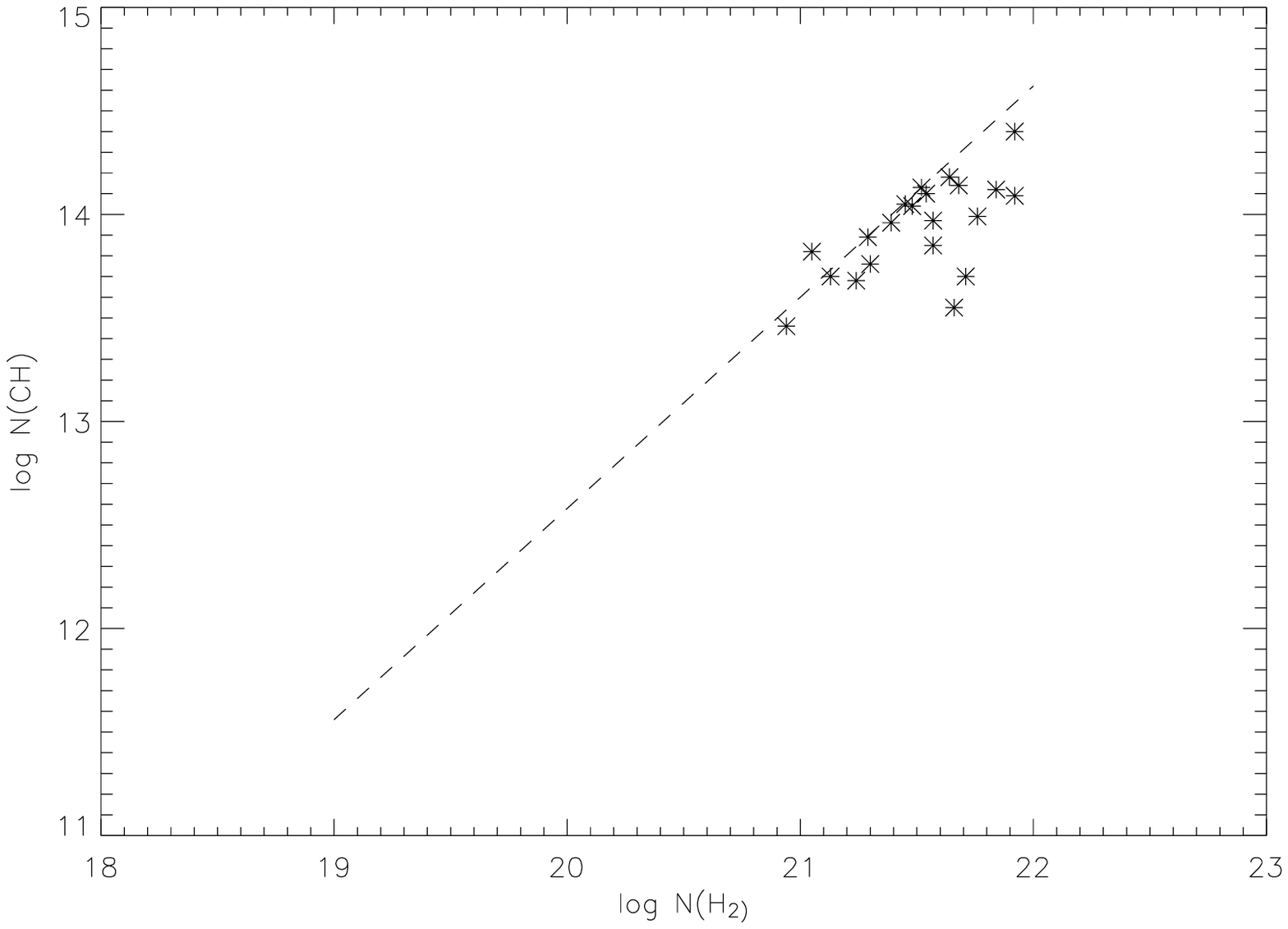}
\caption{
Plot of log N(CH) vs. log N(H$_2$) (column 6 of Table 1 vs. column 5 of Table 2) 
for comparison with Figure 10 of
Mattila (1986).
The dashed line is the least squares fit to the translucent and dark
cloud data presented by Mattila: log N(CH) = 1.02[log N(H$_2$)] - 7.82.
}
\end{figure}

\begin{figure}
\figurenum{5}
\plotone{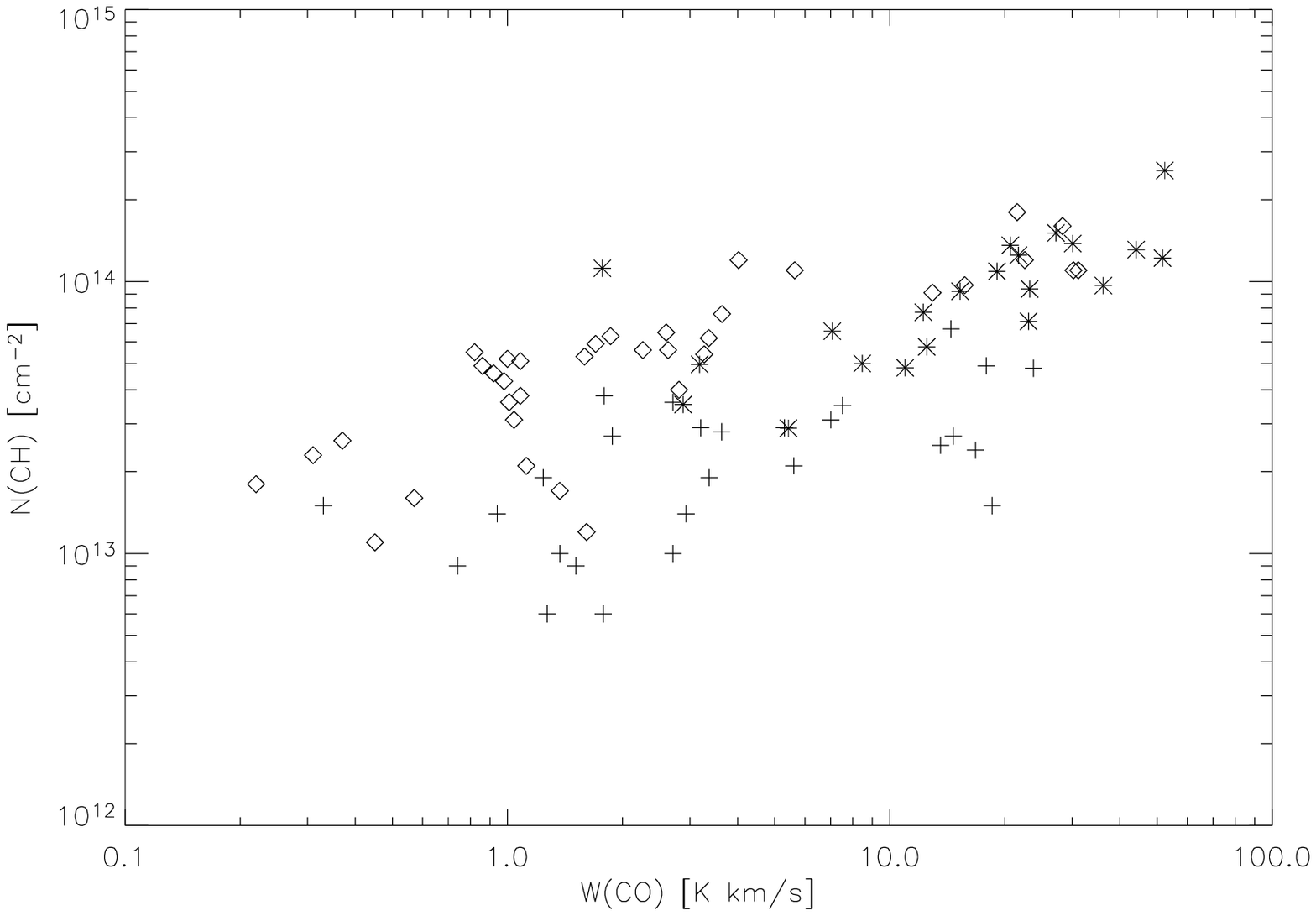}
\caption{
Plot of N(CH) vs. W$_{\rm CO}$ for the data presented in this paper (star symbols),
and the dark and translucent cloud data from Magnani \& Onello (1995) as
categorized by Liszt \& Lucas (2002).  Dark molecular gas is denoted by the
diamond symbol, and the plus signs are for translucent gas.
}
\end{figure}

\begin{figure}
\figurenum{6}
\plotone{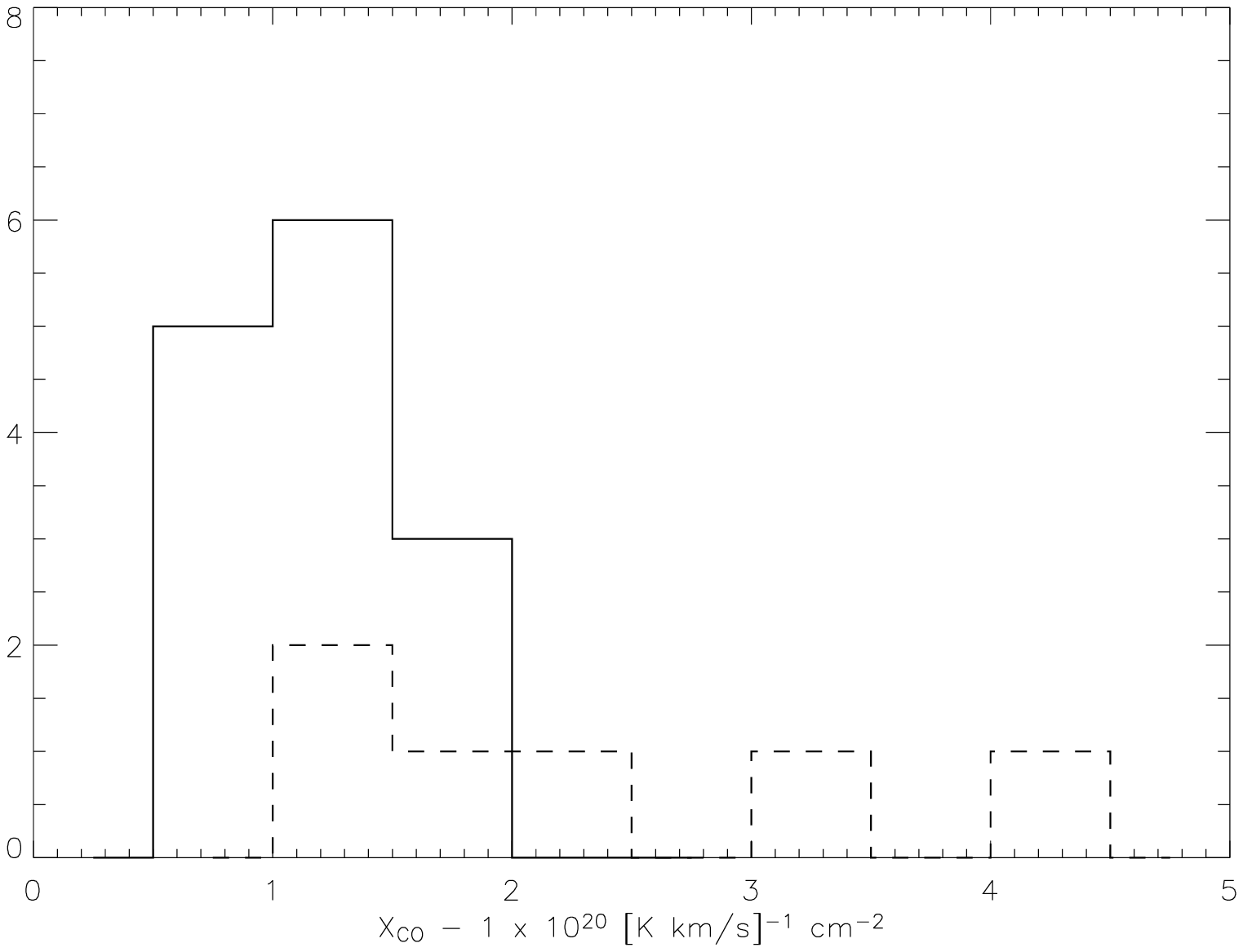}
\caption{
Histogram of the X$_{\rm CO}$ values from Table 2 in units of 1.0 $\times$ 10$^{20}$
[K km s$^{-1}$]$^{-1}$.  The solid lines indicate lines of sight through giant
molecular clouds and the dashed lines indicate dark and/or translucent lines of 
sight (see discussion in \S 5).  Not plotted is a dark/translucent data
point with an X$_{\rm CO}$ value of 14.5.
}
\end{figure}

\begin{figure}
\figurenum{7a}
\plotone{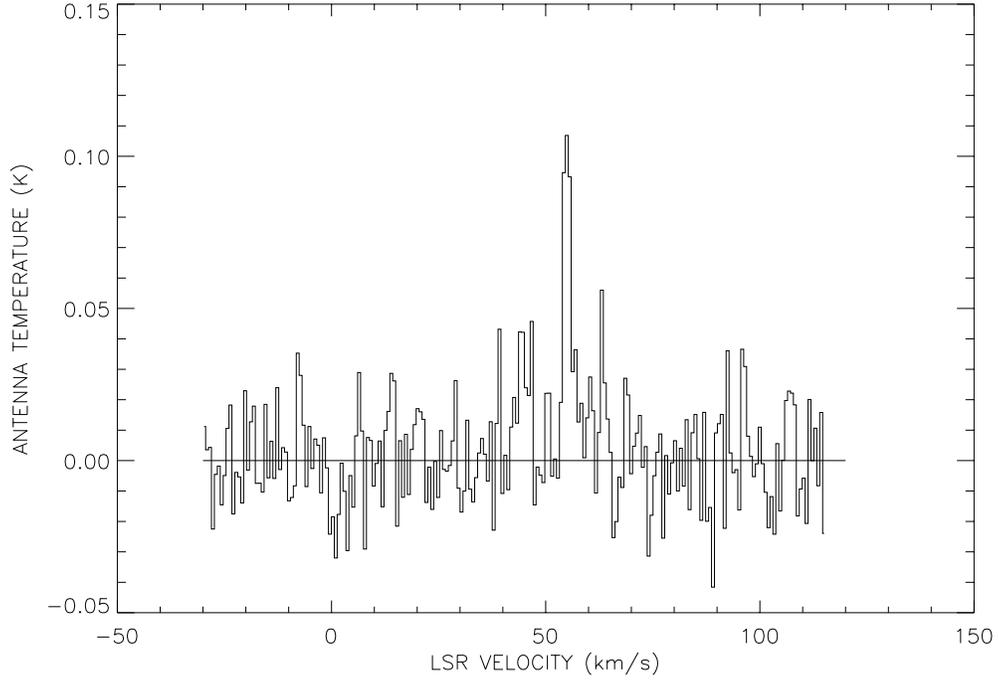}
\caption{
C$^{18}$O(1-0) spectrum of $\ell =$ 50.0$\arcdeg$, $b =$ 0.0$\arcdeg$. The beam size 
is 8.4$\arcmin$ and the velocity resolution is 0.68 km s$^{-1}$ (see \S 2 for more details).
}
\end{figure}

\begin{figure}
\figurenum{7b}
\plotone{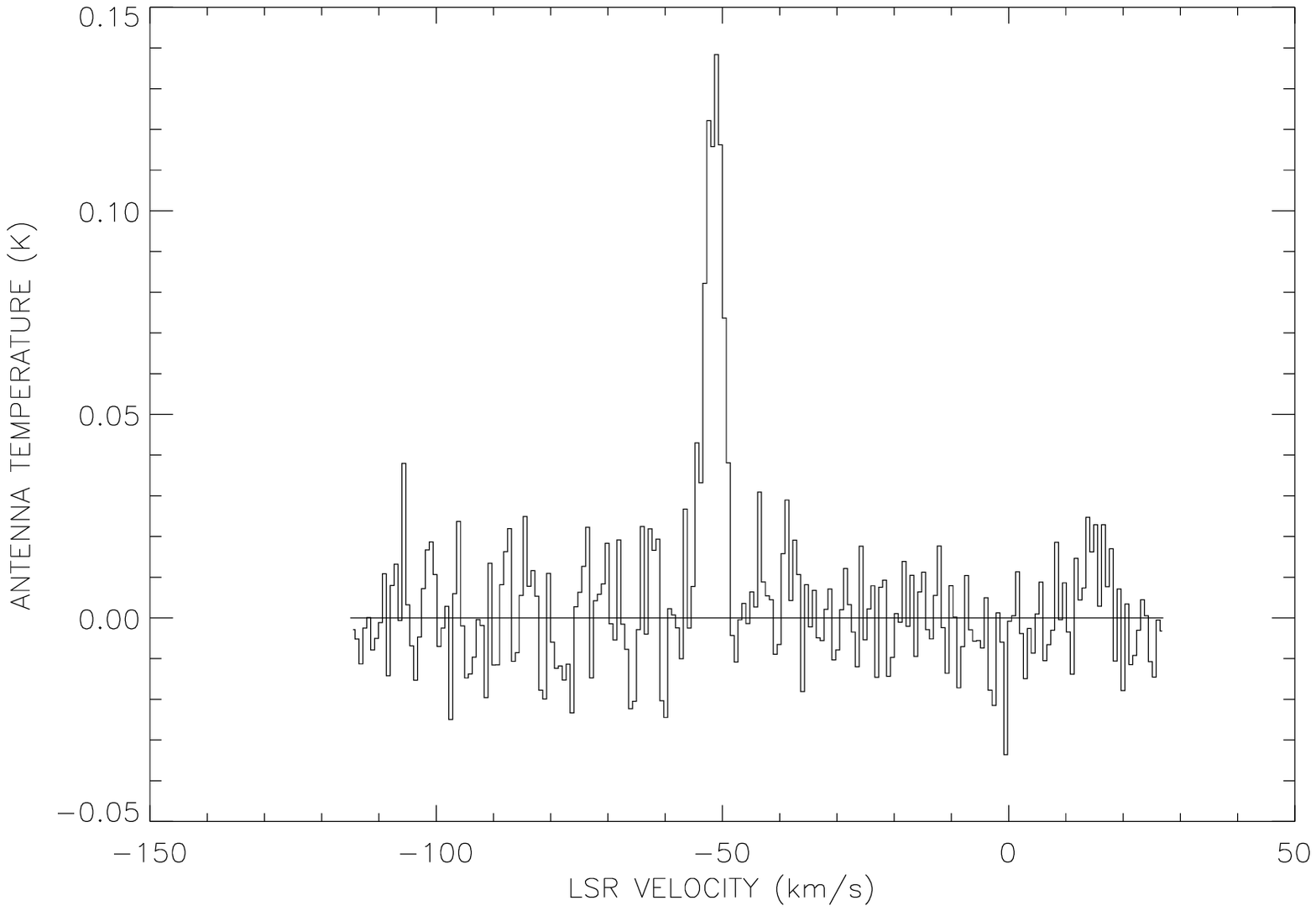}
\caption{
Same as Figure 7a, but for  $\ell =$ 110.0$\arcdeg$, $b =$ 0.0$\arcdeg$.
}
\end{figure}

\end{document}